\documentclass[11pt]{article}
\usepackage{times}
\usepackage{geometry}
\usepackage[utf8]{inputenc}
\usepackage{enumitem,amssymb}
\usepackage{graphicx}
\usepackage{fancyhdr}
\usepackage{aas_macros}
\usepackage{mdframed} 
\usepackage{hyperref}

\usepackage[authoryear]{natbib}
\setcitestyle{authoryear,open={(},close={)}}
\mdfdefinestyle{theoremstyle}{
innertopmargin=\topskip,}
\mdtheorem[style=theoremstyle]{lrptextbox}{}

\begin{document}

%

\begin{center}


{\Large Development Plans for the Atacama Large Millimeter/submillimeter Array (ALMA)}\\
\vspace{4mm}


Christine Wilson$^{1}$\footnote[1]{wilsoncd@mcmaster.ca}, Scott Chapman$^2$, Ruobing Dong$^3$, James di Francesco$^4$, Laura Fissel$^5$, Doug Johnstone$^4$, Helen Kirk$^4$, Brenda Matthews$^{3,4}$, Brian McNamara$^6$, Erik Rosolowsky$^7$, Michael Rupen$^4$, Sarah Sadavoy$^5$, Douglas Scott$^8$, Nienke van der Marel$^4$\\
\emph{\small 1) McMaster University, 2) Dalhousie University, 3) University of Victoria, 4) NRC Herzberg, 5) Queen's University, \\ 6) University of Waterloo, 7) University of Alberta, 8) University of British Columbia}\\
\end{center}

\begin{abstract}
The Atacama Large Millimeter/submillimeter Array (ALMA) was the top-ranked priority for a new ground-based facility in the 2000 Long Range Plan. Ten years later, at the time of LRP2010, ALMA construction was well underway, with first science observations anticipated for 2011. In the past 8 years, ALMA has proved itself to be a high-impact, high-demand observatory, with record numbers of proposals submitted to the past few annual calls (more proposals than are submitted annually to HST) and large numbers of highly cited scientific papers across fields from protoplanetary disks to high-redshift galaxies and quasars. Since Cycle 4, ALMA has also begun to carry out large programs using more than 60 hours of observing time on the 12-m array or more than 200 hours of time on the Atacama Compact Array (ACA).

ALMA’s scientific impact reaches into nearly every area of astronomy. Highlights include the first image of a supermassive black hole in the centre of M87 by the Event Horizon Telescope Collaboration; Canadians led the analysis that extracted the physics, such as black hole mass and spin, from that image. A Canadian-led collaboration has shown that radio galaxies located in clusters and groups can drive molecular gas flows (both inflows and outflows) up to 10s of kpc in altitude. Canadians are also leading innovative studies of proto-stellar and proto-planetary disks, including the first systematic study of their morpohologies and the location of gaps that can signal unseen planets. VERTICO (PI. T. Brown) is the first Canadian-led ALMA large program. VERTICO will map 51 spiral galaxies in the nearby Virgo cluster and use a multi-wavelength approach to quantify the effect of cluster environment on the star-forming molecular gas.

The LRP2010 ALMA white paper laid out 8 specific metrics for ALMA and the Canadian ALMA user community that could be used to judge the success of Canada’s participation in ALMA.  These metrics ranged from publications (number; impact) to collaborations (international; multi-wavelength) to student training and leadership in ALMA operations, as well as the successful completion of the Band 3 (3mm, 100 GHz) receivers and ALMA development projects. All 8 metrics argue for Canada’s involvement in ALMA over the past decade to be judged a success. The successful achievement of these wide-ranging goals argues strongly for Canada’s continuing participation in operating and developing ALMA over the next decade and beyond.

To call out one particular success metric, Canadians are making excellent use of ALMA in training graduate students and postdocs. For example, 12 of 23 Canadian first-author papers published as of June 2018 were led by a graduate student, and a further 4 papers were led by postdocs. As of that date, the Canadian-led paper with the highest number of citations was by a graduate student. The first Canadian-led ALMA large program (VERTICO) is led by postdoc T. Brown at McMaster University.

The ALMA observatory has identified a set of short and medium-term development goals that will keep ALMA at the cutting-edge of astronomy and allow it to continue producing transformational scientific results in future decades. Over the next decade, the focus is on expanding the spectral bandwidth of ALMA by a factor of at least two. This increase in bandwidth requires upgrades to ALMA’s receivers, electronics, and correlator. These increases in bandwidth will reduce the integration time required for a variety of scientific programs, such as blind redshift surveys, spectral scans, and sensitive continuum imaging, by a factor of two. High-resolution spectral scans, for example of proto-planetary disks, will see an increase in speed by a factor of 8 or more. Improvements to the ALMA Archive is another important focus, particularly in the area of applying data mining to large spectral datasets. There are opportunities for Canadian participation and/or leadership in many of these development areas.

Looking forward to the next decade of ALMA operations, our community needs to:
(1) maintain Canadian access to ALMA and our competitiveness in using ALMA; (2) preserve full Canadian funding for our share of ALMA operations; (3) identify components of ALMA development in which Canada can play a significant role, including stimulating expertise in submillimetre instrumentation to capitalize on future opportunities; and (4) keep Canadians fully trained and engaged in ALMA, as new capabilities become available, reaching the widest possible community of potential users.
\end{abstract}

\newpage


\section{Introduction}

The Atacama Large Millimeter/submillimeter Array (ALMA) was the top-ranked priority for a new ground-based facility in the 2000 Long Range Plan. Ten years later, at the time of LRP2010, ALMA construction was well underway, with first science observations anticipated for 2011. In the past 8 years, ALMA has proved itself to be a high-impact, high-demand observatory, with record numbers of proposals submitted to the past few annual calls (more proposals than are submitted annually to the Hubble Space Telescope) and large numbers of highly cited scientific papers across fields from protoplanetary disks to high-redshift galaxies and quasars. Since Cycle 4, ALMA has also begun to carry out large programs using more than 60 hours of observing time on the 12-m array or more than 200 hours of time on the Atacama Compact Array (ACA).

The organization of this White Paper is as follows. In Section~\ref{science}, we give some scientific highlights from research with ALMA since the start of observing in fall 2011. Section~\ref{sec-success} reviews the success criteria for Canadian participation in ALMA that were laid out in the LRP2010 ALMA White Paper. Section~\ref{sec-science_drivers} describes the three new science drivers that have been formulated to guide ALMA development over the next 10--15 years. Section~\ref{sec-development} describes the plans for ALMA development over the short to medium term. Section~\ref{sec-cdn_development} discusses some of the possible Canadian contributions to ALMA development in the next decade. Section~\ref{sec-recommendations} contains our recommendations to the LRP 2020 panel and the responses to the eight specific questions posed by them.

\section{ALMA Science highlights 2011--2019}\label{science}

The scientific potential of the submillimetre waveband, as laid out in previous Canadian planning documents \citep{2010arXiv1008.4159S, 2013arXiv1312.5013W}, has been realised through almost a decade of successful ALMA operations.

\begin{figure}[htbp!]
\includegraphics[width=0.44\textwidth]{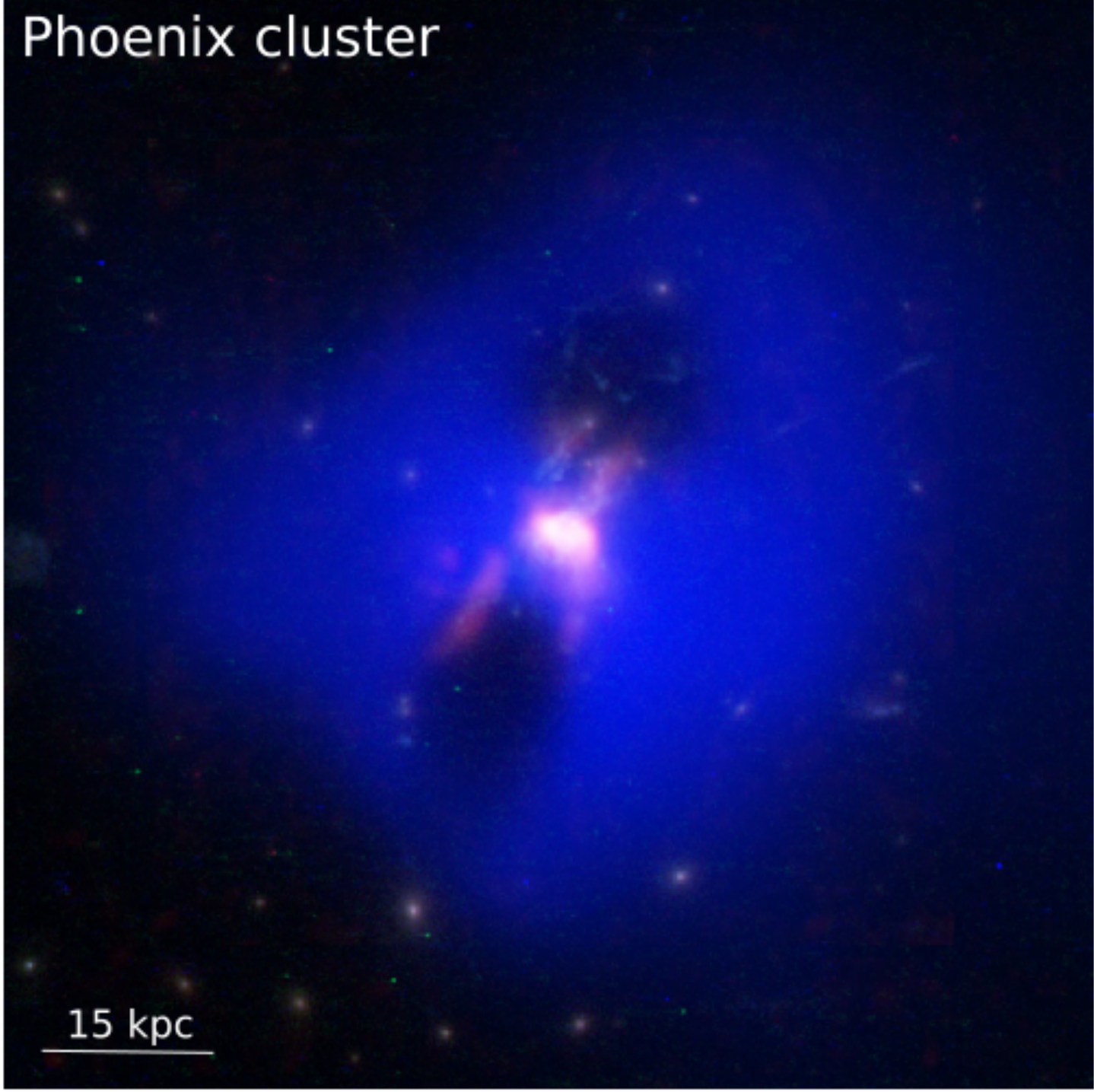}
\includegraphics[trim=0 -0.5cm 0 0, width=0.55\textwidth]{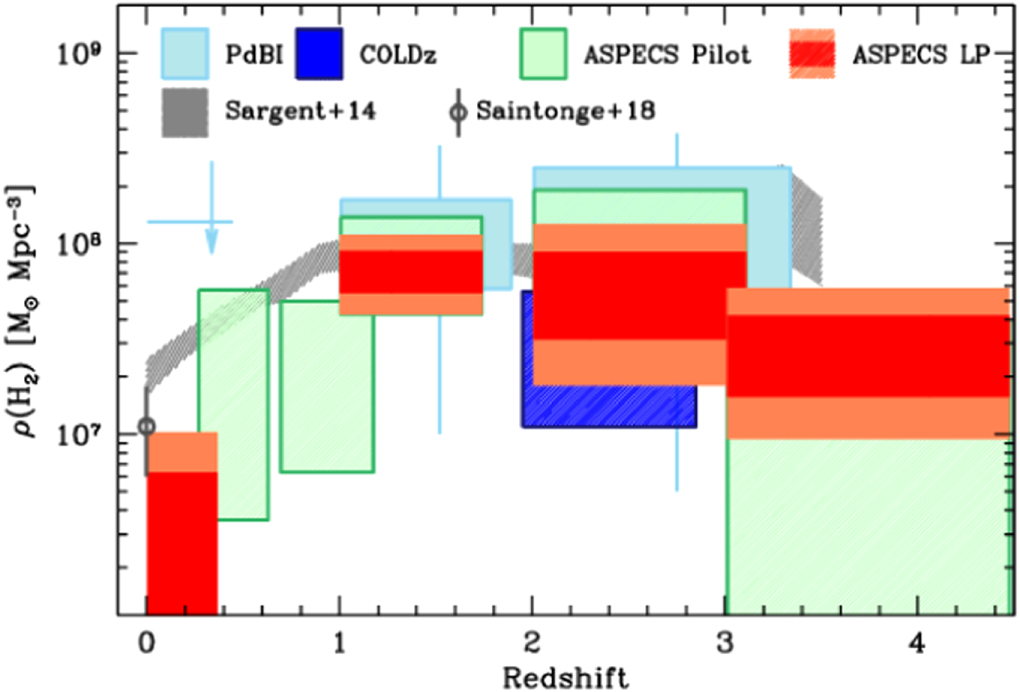}
\includegraphics[trim=0 -1.0cm 0 0, width=0.6\textwidth]{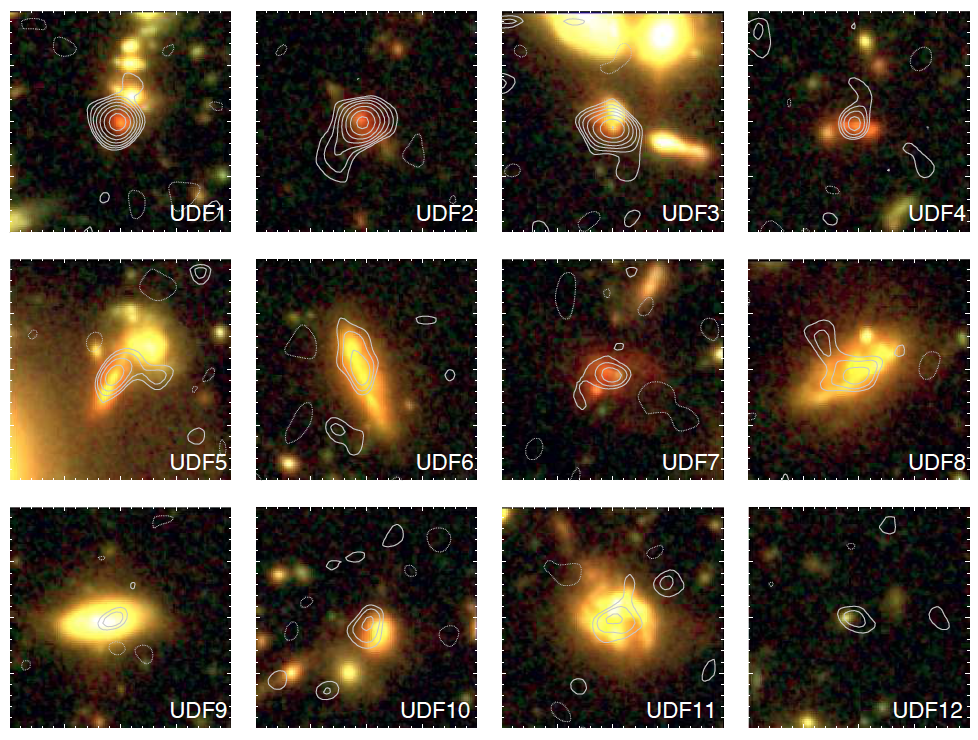}
\hspace{0.25cm}
\includegraphics[width=0.35\textwidth]{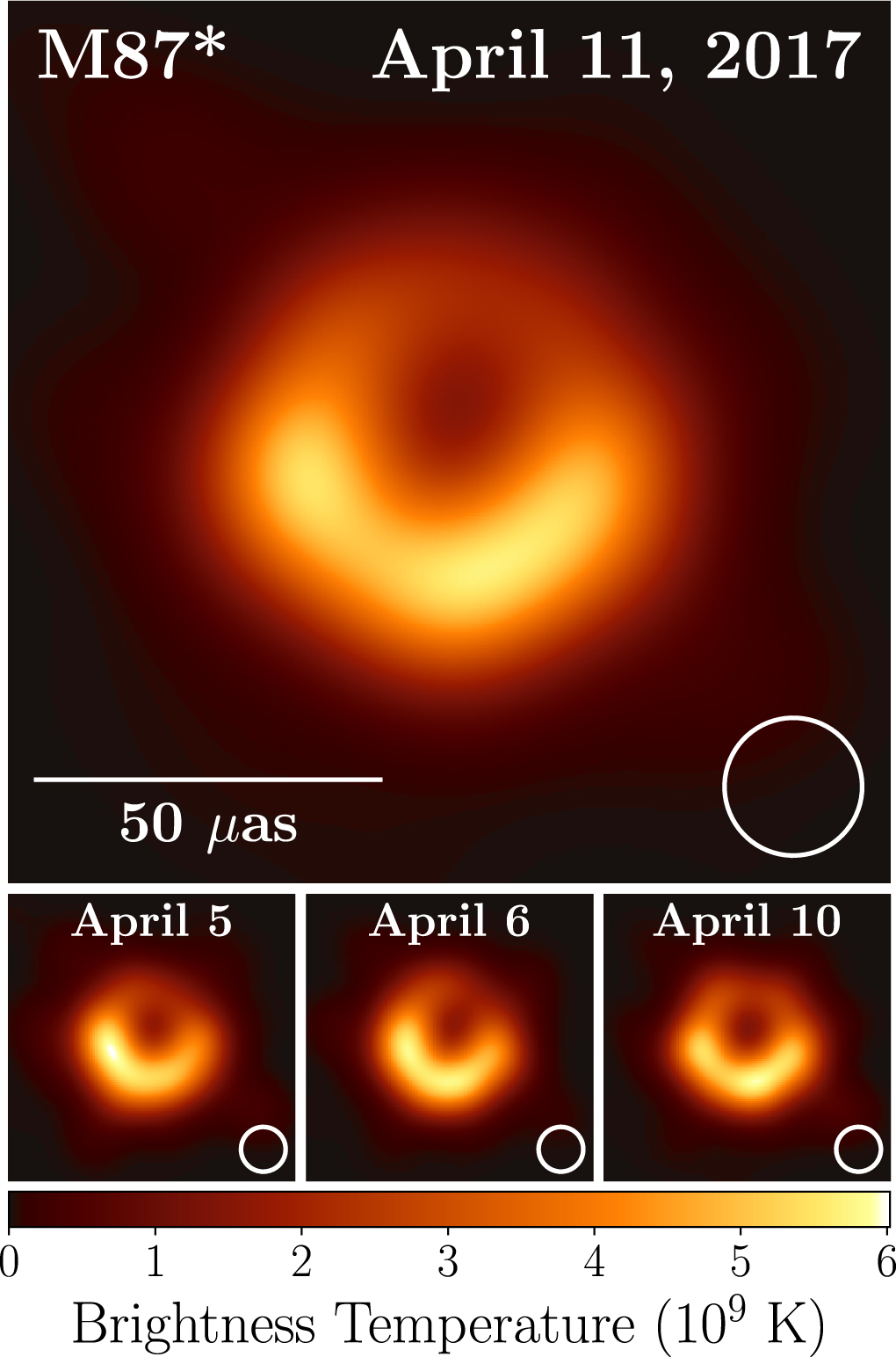}
\caption{A montage illustrating the diversity of extragalactic science targets observed with ALMA. {\bf Top left}: Composite X-ray (blue) and CO image of the central galaxy in the Phoenix
Cluster.  The image shows upward of $10^{10}\,$M$_\odot$ of molecular gas
being lifted out of the galaxy along the edges of rising X-ray bubbles (black).
The X-ray bubbles were inflated by the galaxy’s radio source (not shown).  
Much of the molecular gas may have cooled and condensed out of hot atmospheric 
gas being lifted behind the X-ray bubbles \citep{2017ApJ...836..130R}. 
{\bf Top right}: Redshift evolution of the cosmic molecular gas surface density from ASPECS and other sources \citep{2019ApJ...882..138D}.
{\bf Bottom left}: 12 brightest mm-wave galaxies in the Hubble Ultradeep Field, with ALMA contours shown on top of 3-colour HST images, illustrating the diversity of high-$z$ sources \citep{2017MNRAS.466..861D}. {\bf Bottom right}: EHT image of the black hole in M87 from \citep{2019ApJ...875L...1E}.\label{fig-russell}}
\end{figure}

ALMA was the enabling facility for the first image of a supermassive black hole in the centre of M87 \citep{2019ApJ...875L...1E}. These results and images (Fig.~\ref{fig-russell}) received intensive media coverage across the globe and are without a doubt the single highest-impact science result to come out of ALMA to date.
 Canadians led the analysis that extracted the physics from that image, such as black hole
mass and spin. Avery Broderick 
(U. Waterloo) was on the panel that
presented the results at a media event in Washington, D.C. 
The Event Horizon Telescope (EHT) Collaboration, which includes several Canadians,
has recently been awarded the Breakthrough Prize in Fundamental Physics.


Extremely high-resolution imaging with ALMA has been used to provide evidence for substructure in the lensing halo of SDP.81
\citep{2016ApJ...823...37H}. By applying Bayesian modelling to the uv-data of this strongly lensed source, they find evidence for a dark matter sub-halo of mass around $10^9\,$M$_\odot$ and are able to put constraints on the population of dark matter sub-halos down to masses of $2 \times 10^7\,$M$_\odot$. 

Understanding that galaxies evolve
under the influence of energetic feedback from supermassive black holes represents a
significant advance in our understanding of galaxy evolution.
A collaboration led by Canadian researchers beginning in Cycle 0 has
shown that radio galaxies located in clusters and groups drive molecular
gas flows (inflow/outflow) several to tens of kpc in altitude
\citep[e.g.,][]{2017ApJ...836..130R}. 
 The masses range between $10^9\,$M$_\odot$ to
upward of $10^{10}\,$M$_\odot$, the largest observed among AGN including quasars (Fig.~\ref{fig-russell}).  How radio jets and the X-ray bubbles they
 inflate into their surrounding hot atmospheres are able to lift such large masses is not understood.
Evidence suggests that the feedback mechanism itself stimulates atmospheric
cooling into molecular clouds.  The molecular gas fuels a long-lived feedback loop 
that drives the co-evolution of massive black holes and their host galaxies.  This process may be
occurring at some level in all massive galaxies, indicating a significant scientific advance 
enabled by ALMA.

\begin{figure} [htbp!]
\includegraphics[width=0.42\textwidth]{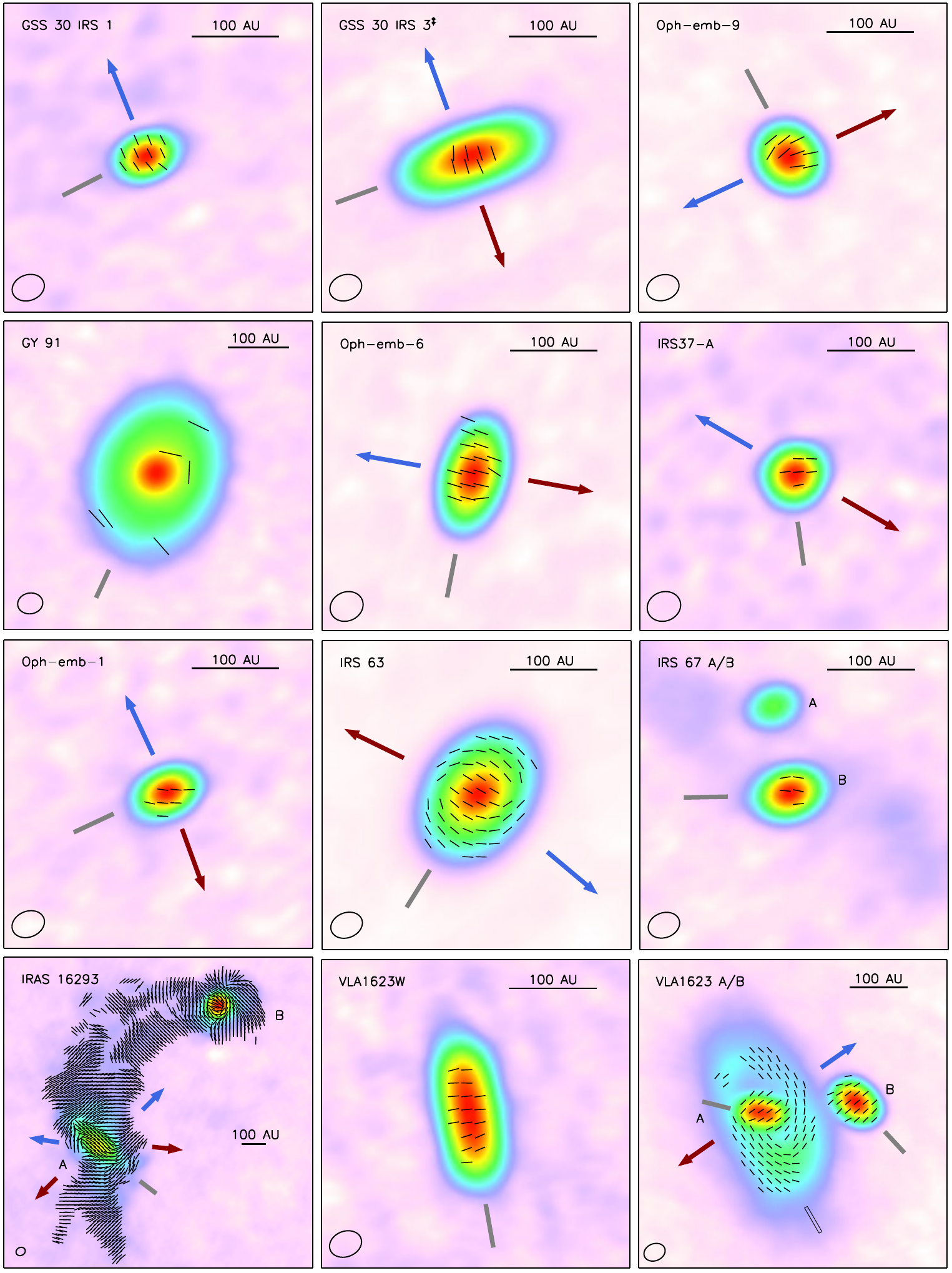}
\includegraphics[width=0.6\textwidth]{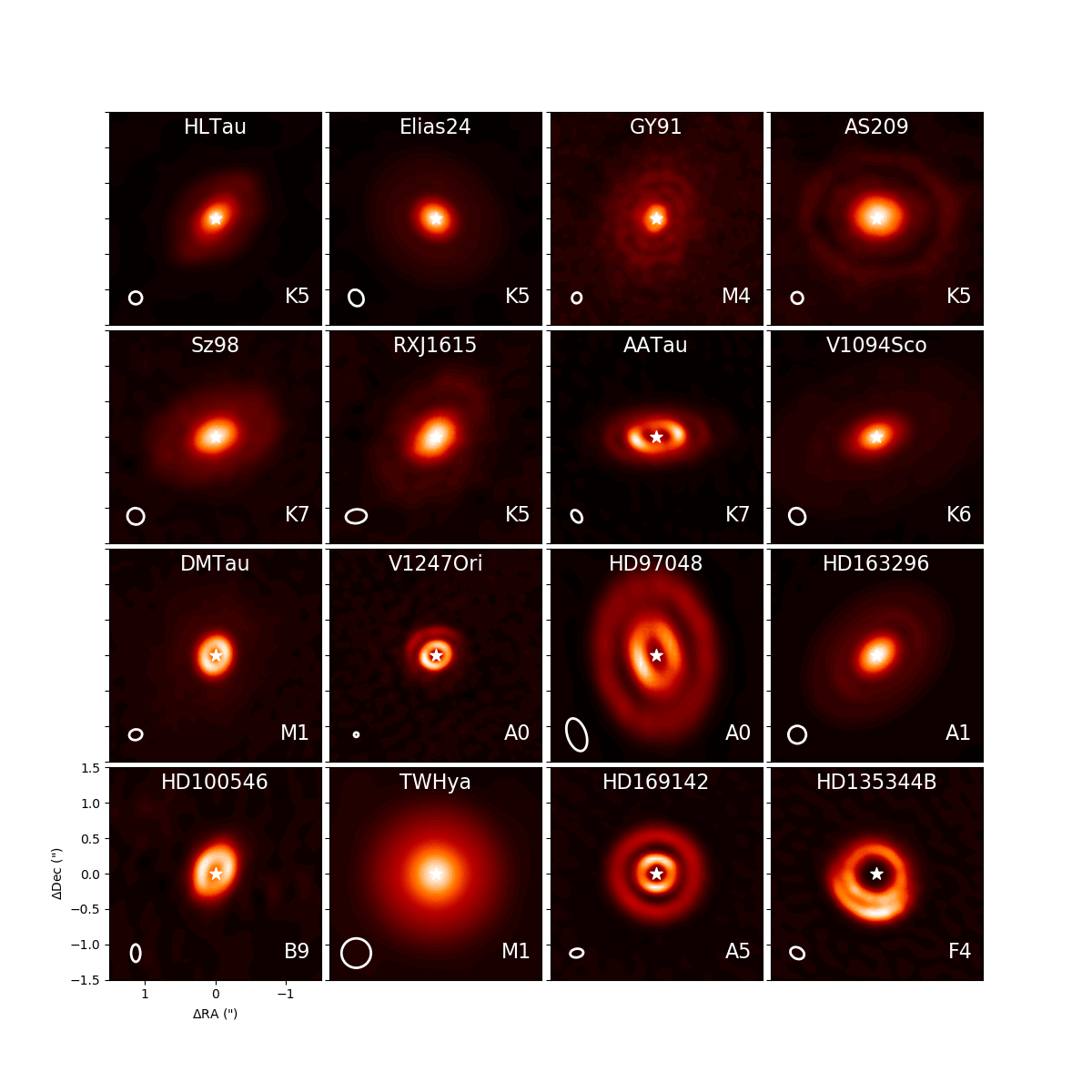}
\caption{A gallery of protostellar and protoplanetary disks observed with ALMA. {\bf Left}: The 14 continuum sources detected in polarized emission in the Ophiuchus molecular cloud by \citet{2019arXiv190902591S}.  Lines denote the normalized polarization pseudo-vectors and background colour is the integrated intensity (Stokes $I$) on a logarithmic scale. {\bf Right}: Survey of 16 proto-planetary disks with rings and gaps from \cite{2019ApJ...872..112V}.
\label{fig-sadavoy}}
\end{figure}

ALMA has revealed a large diversity of structures in protoplanetary disks in both gas and dust, e.g. gaps, rings and asymmetries (Fig.~\ref{fig-sadavoy}), which could be linked to the presence of planets 
\citep[e.g.,][]{2015ApJ...809...93D,2019ApJ...872..112V}. 
Detailed studies of GW Ori (Bi et al., in prep.) have revealed multiple, mis-aligned dust rings that may be produced by disk-star interactions in this triple-star system.
Also, ALMA has uncovered the observational evidence for dust trapping in disks through multi-wavelength observations, a phenomenon to enhance dust growth at the start of planet formation \citep[e.g.,][]{2013Sci...340.1199V}. 
Furthermore, ALMA disk snapshot surveys have mapped the disk mass distributions in all nearby star forming regions 
\citep[e.g.,][]{2018ApJ...859...21A}. 
Disk masses appear to be too low to form exoplanetary systems at $\sim$2 Myr, indicating that either disk masses are severely underestimated or planet formation is already close to finished at this stage \citep[e.g.,][]{2018A&A...618L...3M}.
ALMA observations of the kinematics of protostellar disks have shown that these disks can have extremely low levels of turbulence, with turbulent broadening at a level of less than 10\% of the sound speed \citep{2018ApJ...856..117F}. These results limit the magnetic viscosity parameter $\alpha < 0.007$ and are driving a major re-examination of our thinking on turbulence in disks, which can have dramatic effects on models of planet formation \citep[e.g.][]{2019ApJ...875...43K}. 

Polarization observations of protoplanetary disks with ALMA, the only facility that can resolve their polarized emission, are also producing surprises. The changing polarization morphology as a function of wavelength observed in HL Tau points to the polarization being produced by dust self-scattering processes at shorter wavelengths and possibly by grains aligned via radiation anisotropy at longer wavelengths \citep[e.g.][and references therein]{2017ApJ...851...55S}, rather than the classical picture of dust grains aligned via magnetic fields. These observations bring into question our ability to use polarization to trace magnetic field morphology, at least in very massive disks like HL Tau. \citet{2019arXiv190902591S} have used ALMA to carry out a complete polarization survey with 35 AU resolution of all the deeply embedded protostars in the nearby Ophiuchus molecular cloud (Fig.~\ref{fig-sadavoy}). They find that the majority of these lower-mass disks  have morphologies consistent with dust self-scattering in optically thick disks (Fig.~\ref{fig-sadavoy}).

The surveys carried out with ALMA large proposals have also produced interesting results. One of the first two surveys, ASPECS, the ALMA Spectral line survey in the UDF (the Hubble Ultra-Deep Field, used 
ALMA to conduct the first blind CO survey of the high-redshift universe \citep[e.g.,][]{2019ApJ...882..138D}. ALMA has helped to demonstrate that the rise and fall in the cosmic star formation rate as a function of redshift is primarily driven by a similar rise and fall in the molecular gas content, the fuel for star formation (Fig.~\ref{fig-russell}).
The second survey, DSHARP, Disk Substructures at High Angular Resolution Project, is a survey of 20 nearby protostellar disks with 5-AU resolution \citep{2018ApJ...869L..41A} in both CO and continuum emission.


More recent ALMA large projects include an ambitious survey of 100,000 giant molecular clouds across 70 nearby spiral galaxies (PHANGS, PI E. Schinnerer), a complete astrochemical survey of the nearest starburst galaxy NGC 253 (ALCHEMI, PI. S. Martin), a systematic investigation of [{\sc Cii}] in the early Universe (ALPINE, PI. O Le F\`evre), and a systematic study of the conditions in molecular clouds that set the stellar initial mass function (PI F. Motte). The PHANGS survey is providing unmatched data on the structure of the star-forming interstellar medium in the local Universe \citep{2018ApJ...860..172S}. The ALPINE survey has recently discovered a rare triple merger at $z=4.56$ \citep{2019arXiv190807777J}, indicating a major growth phase which will likely produce a single massive galaxy by $z\sim 2.5$.
In Cycle 7, the first Canadian-led large program (VERTICO, PI. T. Brown) will survey CO emission in Virgo spirals with the ACA to probe the effect of the cluster environment on the molecular gas.

\section{Has
Canadian participation in ALMA been a success?}
\label{sec-success}

The ALMA White Paper submitted to LRP2010 
\citep{wilson2010}
laid out a number of specific accomplishments by ALMA and the Canadian ALMA user community that would likely lead to our community viewing its participation in ALMA as a success. Ten years on, it is enlightening to review this list and see how we did. All publications statistics are from C. Wilson's ALMA overview at the June 2018 CASCA meeting.

\begin{enumerate}

\item {\it Significant numbers of Canadian first author papers are based on
  ALMA data or theoretical
interpretations of ALMA data}


As of June 2018, approximately 2.2\% of ALMA papers (24 papers total) had a Canadian-based researcher as first author. This fraction is very close to our financial share of global ALMA operations. An additional 11.6\% of ALMA papers (approximately 130 papers in total) had a Canadian-based researcher as a co-author of the paper.

\item {\it Canadian first author papers based on
  ALMA data or theoretical
interpretations of ALMA data have a high impact}


As of June 2018, four of the 70 most highly cited ALMA papers (5.7\%) had a Canadian-based researcher as first author. 
\citet{2013ApJ...767..132H}
described ALMA observations of strongly lensed dusty galaxies discovered by the South Pole Telescope. \citet{2013ApJ...770...13W,2015ApJ...807..180W} measured the star-formation rates and dynamical masses of galaxies hosting supermassive black holes at a redshift $z=6$. \citet{2014ApJ...785...44M} presented high-resolution observations of a flow of $10^{10}\,$M$_\odot$ of molecular gas in the cooling flow of the brightest cluster galaxy in the Abell 1835 cluster.

\item {\it Canadians are playing important roles in some of the very
  highest profile ALMA papers coming from large international teams}
  
  
  Canadian research with ALMA spans a wide range of topics. Three examples are given here.
  
 (1)  Canadian researchers have played leading roles in many of the results coming from ALMA observations of high-redshift sources identified by the South Pole Telescope \citep[e.g.,][]{2013ApJ...767..132H}. \citet{2018Natur.556..469M} revealed a massive core for a cluster of galaxies at a redshift of 4.3 that could be building one of the most massive structures in the Universe.
  
  (2) Canadians are playing leadership roles in two of the 14 ALMA Large Programs approved as of July 2019. The PHANGS survey is mapping 100,000 giant molecular clouds in 70 nearby galaxies (co-PI E. Rosolowsky, U. Alberta). The Canadian-led VERTICO survey is mapping molecular gas on 0.5-kpc scales in all the spiral galaxies in the Virgo Cluster (PI T. Brown, co-PI C. Wilson, McMaster University).
  
  (3) Canadians are also making significant contributions in studying protoplanetary disks. For example, \citet{2019ApJ...872..112V} have published a survey of 16 disks showing ring-like structures. This paper is the first systematic study of morphologies and gap locations and has already collected 24 citations in less than a year.

\item {\it Significant numbers of Canadian astronomers who do not have
  radio astronomy as their primary background are involved in
  ALMA science, either as a primary user of data or as a significant
  collaborator}
  
  
  Canadians at NRC and the University of Calgary led the development of the ``ALMA Primer.'' This ${>}20$-page overview of ALMA's capabilities was first introduced in Cycle 0 and has been updated at each subsequent Cycle. By describing ALMA capabilities in simpler terms and including a variety of example projects and observations, the ALMA Primer has played a significant role in making ALMA more accessible to new users, be they students or more senior researchers without significant millimetre or interferometry experience. 
  
  One example of diverse Canadian participation is the VERTICO large program approved for ALMA Cycle 7. VERTICO has a total of 26 co-Is and includes 5 out of 9 Canadian co-Is whose expertise is in optical astronomy or numerical simulations of galaxies.

\item {\it Canadian graduate students are participating or leading ALMA
  papers and using ALMA data in their theses}
  
  
  As of June 2018, 12 of 23 Canadian first-author papers were led by graduate students and a further four were led by postdocs. Indeed, the highest cited Canadian-led paper at that time was by a graduate student \citep{2013ApJ...767..132H}.
  The 2019 Plaskett Medal winner, Alexandra Tetarenko, used data from ALMA (and many other telescopes!) in her award-winning Ph.D. thesis. An incomplete list of other Canadian Ph.D. theses to use ALMA data includes K. Sliwa (McMaster), A. Vantyghem (Waterloo), Y. Hezaveh (McGill), and J. White (UBC). Ongoing Ph.D. theses using ALMA work include
  R. Hill (UBC), L. Francis (UVic), J. Bi (UVic), A. Bemis (McMaster), N. Brunetti (McMaster), and P. Tamhane (Waterloo).

Postdocs working in Canada have led high-impact programs and papers. For example, T. Brown (McMaster) is the PI of the first Canadian-led ALMA large program (VERTICO), which will measure the effect of environment on the molecular gas in spiral galaxies in the VIRGO cluster. H. Russell (Waterloo) played a major role in the studies of cooling-flows around brightest cluster galaxies \citep{2017ApJ...836..130R}. R. Mann (NRC) carried out a detailed study of five proplyds in Orion, revealing rapid dissipation of disks that would inhibit planet formation in UV-dominated environments  \citep{2014ApJ...784...82M}. R. Friesen (Dunlap Institute) identified what may be the first hydrostatic core in high-resolution observations of the Ophiuchus star forming region \citep{2014ApJ...797...27F}.

\item {\it The Band-3 receivers are delivered on time, meet or exceed ALMA
  specifications, and are
being used for good science by international ALMA community}


The Band 3 receivers were one of the four receiver bands available in ALMA first Early Science call for proposals (Cycle 0). They continue to be in high demand, second only to Band~6 in the amount of time requested or awarded per Band across all 66 ALMA antennas.

\item {\it Canada is leading or playing a major role in an interesting ALMA
  development project such as the Band-1 receivers}
  
  
  ALMA Development Projects are meant to provide physical or software deliverables to be incorporated in the ALMA observatory. Across the observatory, examples of Development Projects range from the construction of the Band-5 receivers to the phasing hardware and software that enabled ALMA to join the Event Horizon Telescope.
  
  Two ALMA Development Projects have been led by Canadians. One was ``CARTA: The Next Generation ALMA Viewer'' led by E. Rosolowsky (UBC/U.Alberta), which was designed to replace the existing viewer in CASA and has been released to the user community (\url{https://cartavis.github.io}). The second was ``Band-3 cold cartridge assembly magnet and heater installation for deflux operation'' led by L. Knee (NRC) . Canada has also been participating in the Band-1 receiver development project led by Taiwan.
  
ALMA Development Studies are typically one-year programs with relatively modest funding, that may eventually lead to a Development Project. As an example, the Canadian-led CARTA viewer initially began as a development study. Canadians have led six of the 29 development studies awarded funding. In the most recent call in 2019, two ALMA development studies were awarded to Canadian teams from NRC. One program is ``High level design and integration of NRC TALON-based correlator for increased channels, bandwidth, and baselines.'' The other program is ``ARCADE: ALMA Reduction in the CANFAR Data Environment'' (see Section~\ref{sec-cdn_development} for more on ARCACE).

\item {\it Canadians are playing a leadership role in some aspect of ALMA
operations in North America or internationally}


The current ALMA Director, Sean Dougherty, is a Canadian. Eduardo Hardy was the legal representative of AUI (the U.S.-based not-for-profit corporation that holds the NSF grant to operate ALMA and NRAO) in Chile for many years and remains their senior advisor in Chile.  Rachel Friesen (currently at UofT) was a staff member at the North American ALMA Science Center. Christine Wilson served on the search committees for the second and fourth ALMA Directors. Jim Hesser was a long-serving member of the ALMA Board and chaired its Budget Committee.  James Di Francesco is currently the Canadian member of the ALMA Board.  There has always been a Canadian as one of the North-American representatives on the ALMA Science Advisory Committee (currently C. Wilson).  

\end{enumerate}

Based on these metrics laid out a decade ago,  the Canadian astronomical community should view 
our participation in ALMA as a
success. {\bf The successful achievement of these wide-ranging goals argues strongly for Canada's continuing participation in operating and developing ALMA over the next decade and beyond.}




\section{New fundamental science drivers for ALMA}
\label{sec-science_drivers}


In consultation with the ALMA user community, a new set of top-level science goals has been developed to guide ALMA development over the next decade
\citep{2019arXiv190202856C}.
These new goals are as follows:

\begin{itemize}
    \item {\bf Origins of galaxies:} Trace the cosmic evolution of key elements from the first galaxies ($z>10$) through the peak
of star formation ($z=2–4$) by detecting their cooling lines, both atomic ([{\sc Cii}], [{\sc Oiii}]) and
molecular (CO), and dust continuum, at a rate of 1--2 galaxies per hour.
\item {\bf Origins of chemical complexity:} Trace the evolution from simple to complex organic molecules through the process of star
and planet formation down to solar system scales (10--100\,au) by performing full-band
frequency scans at a rate of 2--4 protostars per day.
\item {\bf Origins of planets:} Image protoplanetary disks in nearby ($d\,{<}\,150\,$pc) star-formation regions to resolve the Earth-forming zone ($\sim$1\,au) in the dust continuum at wavelengths shorter than 1\,mm, enabling
detection of the tidal gaps and inner holes created by planets undergoing formation.
\end{itemize}
Achieving these goals will require a set of ambitious upgrades to ALMA over the next 10--15 years. These upgrades will keep ALMA at the cutting edge of astronomy and allow it to continue producing transformational scientific results in future decades.

\section{ALMA
development in the next decade and beyond}
\label{sec-development}

The ALMA2030 report was 
submitted to the ALMA Board in March 2015. In it, four development paths were recommended based on their long-term scientific potential:

\begin{itemize}
    \item improvements to the ALMA archive to achieve gains in usability and increase the impact of the observatory;
    \item larger bandwidths and improved receiver sensitivity to achieve gains in speed;
    \item longer baselines to enable qualitatively new science, and;
    \item increased mapping speed to enable more efficient wide-field imaging.
\end{itemize}
The ALMA Development Working Group subsequently divided these four paths into short-term and medium-term development goals \citep{2019arXiv190202856C}.

\subsection{Near-term
ALMA development priorities}

The current ALMA development priorities are to: 
\begin{itemize}
    \item broaden the receiver IF bandwidth by at least a factor of 2; 
    \item upgrade the digitizers and digital processing to allow for larger bandwidth, and; 
    \item upgrade the correlator to process these larger bandwidths with high spectral resolution.
\end{itemize}
These developments will significantly reduce the time required for a wide range of scientific applications. For example, the time required would be reduced by a factor of 2 for 
blind redshift surveys, spectral scans, and deep continuum surveys, while the time required for high spectral-resolution spectral scans (for example in low-mass protostars or evolved stellar envelopes) would be reduced by factors of 8--16.

In terms of receiver upgrades, the priority is: (1) intermediate frequencies 200--425\,GHz; (2) lower frequences ${<}\,$200\,GHz; and
(3) higher frequencies ${>}\,$425\,GHz. This frequency prioritization was selected to have the most direct impact on enabling the new science drivers. For more details, see \citet{2019arXiv190202856C}.

\subsection{Archive development}

In addition, the Working Group recommended that a committee be tasked with prioritizing the archive capabilities that will be required to facilitate increased scientific exploitation 
of the ALMA archive over the next decade. Particular attention will need to be paid towards facilitating data mining of ALMA's spectral products, especially in light of the planned receiver upgrades. The Canadian-led ARCADE project has the potential to be a useful contribution to archive capabilities (see Section~\ref{sec-cdn_development}).

\subsection{Medium-term development}

The medium-term opportunities include: extended baselines with the addition of at least six additional antennas to ensure the minimum required $uv$ coverage on the longest baselines; focal-plane arrays to increase the mapping speed; and additional 12-m antennas to enhance the sensitivity. All these opportunities were recommended for further development studies of their scientific, technical, and logistical potential and scope.

A large (25--50m) single-dish telescope was noted to provide strong scientific synergies with ALMA, but was thought to be outside the scope of the current ALMA project.    

\section{Potential Canadian contributions to ALMA development 2020--2030}
\label{sec-cdn_development}

There are a variety of paths open to Canada to contribute to ALMA development. For example, the ARCADE study could lead to a significant contribution relevant to archival research. 
ARCADE is a new initiative to make ALMA data more accessible to Canadian astronomers, which aims to provide a virtual computing environment for any researcher’s ALMA data needs (archival projects as well as PI science).  Within the ARCADE environment, users will have access to the large amounts of RAM and storage space needed for processing ALMA data, which may be challenging to obtain at their home institutions.  Researchers will be able to run their preferred version of CASA (note that older versions of CASA are required for re-calibrating archival data sets), the first step needed for analysis, since only raw data is provided in the ALMA archive.  At present, ARCADE is in an early prototyping stage.  The NRC’s Millimetre Astronomy Group and the Canadian Astronomical Data Centre, in collaboration with C.~Wilson’s group at McMaster University, were recently awarded a one year ALMA North American Development Study Proposal for ARCADE, which will allow the system to be further developed.  Specifically, issues related to scalability in a virtual environment (processing resources and storage space) to allow for multiple parallel users will be investigated by CADC.  Meanwhile, the Millimetre Astronomy Group (NRC) and McMaster will undertake significant external user testing to improve the functionality of the system and identify any additional software needs (e.g., independent but CASA-affiliated software such as ADMIT).  Over the course of the next decade, the goal is for ARCADE to be offered to astronomers across Canada, where it can reduce the hardware and software access hurdles that may discourage analysis of ALMA data by experienced and new users alike.
  
Another area where Canada could take the lead would be the upgrade of the Band-3 receivers to wider bandwidth.
Canadians could also take the lead on studies of a particular medium-term development, such as focal-plane arrays. Given our experience with correlator development for the JCMT and the Karl. G. Jansky Very Large Array, it is possible that Canada could play a significant role in some aspects of the upgrade to the ALMA correlator.

Contributing to ALMA development would bring in ``new'' money from the ALMA development funding stream, but could also involve significant in-kind contributions in time and effort 
by scientists at NRC and Canadian universities.

\section{Recommendations}
\label{sec-recommendations}

Canada is a partner in ALMA, and the observatory is being successfully exploited by Canadians. It is important, however, to continue to build on this success. As we move forward, we need to:
\begin{itemize}
    \item maintain Canadian access to ALMA and our competitiveness in using ALMA;
    \item preserve full Canadian funding for our share of ALMA operations;
    \item identify components of ALMA development in which Canada can play a significant role, including stimulating expertise in submillimetre instrumentation to capitalize on future opportunities; and
    \item keep Canadians fully trained and engaged in ALMA, as new capabilities become available, reaching the widest possible community of potential users.
\end{itemize}
Over the past decade, the successful achievement of the wide-ranging goals laid out in the LRP 2010 white paper \citep{wilson2010} 
argues strongly for Canada's continuing participation in operating and developing ALMA over the next decade and beyond.

\bigskip
\bigskip




\begin{lrptextbox}[How does the proposed initiative result in fundamental or transformational advances in our understanding of the Universe?]


In terms of sensitivity, speed, and angular resolution, ALMA is the most powerful (sub-)millimetre interferometer in the world and is already delivering transformational science (see Section~\ref{science}). The ALMA upgrades planned for the next 10--15 years will ensure that ALMA will continue to enable fundamental scientific advances for decades to come (see Section~\ref{sec-development}).

\end{lrptextbox}


\begin{lrptextbox}[What are the main scientific risks and how will they be mitigated?]


There are no significant scientific risks to ALMA development upgrades. However, we will need to ensure that the community has the resources to deal with the (already) large data volumes produced by ALMA. The ARCADE development project (Section~\ref{sec-cdn_development}) is an interesting step in this direction.

\end{lrptextbox}

\clearpage

\begin{lrptextbox}[Is there the expectation of and capacity for Canadian scientific, technical or strategic leadership?] 


Canadians have already demonstrated significant scientific leadership with ALMA (see Sections~\ref{science} and~\ref{sec-success}). The first Canadian-led large proposal (VERTICO, PI. T. Brown) is only the most recent example of our successful scientific leadership. There will be opportunities for Canadian technical and strategic leadership in specific areas of ALMA development over the next 10--15 years(see Section~\ref{sec-cdn_development}).

\end{lrptextbox}

\begin{lrptextbox}[Is there support from, involvement from, and coordination within the relevant Canadian community and more broadly?] 


ALMA is clearly the premier instrument for high-resolution (sub-)millimetre-wave astronomy. The high demand for time on ALMA by Canadians and the increasing number of papers and projects led by Canadians (Sections ~\ref{sec-science_drivers} and~\ref{sec-success})
are signs of our community's involvement, as are the large number of development studies and programs led by Canadians (see Section~\ref{sec-cdn_development}).

\end{lrptextbox}

\begin{lrptextbox}[Will this program position Canadian astronomy for future opportunities and returns in 2020--2030 or beyond 2030?] 

Yes: for example, in scientific results (Section~\ref{science}), student training (Section~\ref{sec-success}), and technical and software development (Section~\ref{sec-cdn_development}).

\end{lrptextbox}

\begin{lrptextbox}[In what ways is the cost-benefit ratio, including existing investments and future operating costs, favourable?] 


As an observatory, ALMA has completed construction and is into steady-date operations. Its annual operating costs are on the order of \$80M/yr (USD) over the three regional partners. Spending on the order of \$80M (shared by all ALMA partners over 10 years) on development work that will produce a more powerful ALMA is an extremely cost-efficient investment. Canada's contribution to ALMA development is part of our share of operating costs. We can recover some of that funding back to Canada by participation and/or leadership of one or more ALMA development programs.

\end{lrptextbox}


\begin{lrptextbox}[What are the main programmatic risks
and how will they be mitigated?] 

International co-ordination of ALMA Development programs could be improved, a point which has been highlighted several times by the ASAC in its reports to the ALMA Board. Without better co-ordination of development efforts across the ALMA community, there is the real risk that these planned improvements to ALMA do not happen in a timely manner.

Another possible risk is that lead partners may turn their attention (and funding) to new projects, such as ESO's ELT, NRAO's ngVLA, or a new NAOJ project. Downward pressure on the ALMA operating budget is also a risk for all partners.

\end{lrptextbox}

\clearpage

\begin{lrptextbox}[Does the proposed initiative offer specific tangible benefits to Canadians, including but not limited to interdisciplinary research, industry opportunities, HQP training,
EDI,
outreach or education?] 

ALMA is having a major impact on HQP training in Canada. This is indicated by the large number of Canadian-led papers for which a student or postdoc is the first author and in the Ph.D. theses that use significant data from ALMA (see Section~\ref{sec-success}). ALMA is in the process of becoming a partner in the CREATE program New Technologies for Canadian Observatories (NTCO, PI. K. Venn) and will soon welcome its first graduate intern from Canada, bringing an additional dimension to graduate student training.

Within Canada, the ALMA user community has a higher proportion of women than average, giving it a role in EDI. For example, in ALMA papers published to June 2018, 2/8 students, 4/4 postdocs, and 2/4 faculty/staff (so 50\% in total) who were first authors on one or more ALMA papers were women.

ALMA results also play a significant role in public outreach, for example in the press coverage of the recent black-hole image from the EHT team, or in public talks, such as C. Wilson's keynote address at Starfest 2016, the largest star party in North America.

\end{lrptextbox}


\end{document}